\documentclass[article,floatfix,reprint,twocolumn,superscriptaddress]{revtex4-2}
\usepackage{xcolor} 
\usepackage{graphicx}
\usepackage[hidelinks]{hyperref}
\usepackage{amssymb}
\usepackage[utf8]{inputenc}
\usepackage{amsmath}
\usepackage{ulem}
\usepackage{bbold}

\usepackage{tikz}
\usetikzlibrary{patterns}
\usepackage{xcolor}
\usepackage{graphicx}

\DeclareSymbolFont{rsfs}{U}{rsfs}{m}{n}
\DeclareSymbolFontAlphabet{\mathscrsfs}{rsfs}

\usepackage{braket}

\makeatletter
\newcommand*{\rom}[1]{\expandafter\@slowromancap\romannumeral #1@}
\makeatother

\newcommand{\abs}[1]{\left|#1 \right|}

\newcommand{\alp}{{\cal A}}

\definecolor{vastkust}{RGB}{0, 120, 10} 
\hypersetup{
  colorlinks,
  citecolor=vastkust,
  linkcolor=vastkust,
  urlcolor=vastkust}
  
\begin{document}
\title{Quantifying non-Markovianity in magnetization dynamics \\ via entropy production rates}

\author{Felix\ Hartmann}
\email{hartmann3@uni-potsdam.de}
\affiliation{University of Potsdam, Institute of Physics and Astronomy, Karl--Liebknecht--Str. 24--25, 14476 Potsdam, Germany}

\author{Finja\ Tietjen}
\email{finja.tietjen@chalmers.se}
\affiliation{Department of Physics, Chalmers University of Technology, 412 96 G\"oteborg, Sweden}

\author{R. Matthias\ Geilhufe}
\affiliation{Department of Physics, Chalmers University of Technology, 412 96 G\"oteborg, Sweden}

\author{Janet\ Anders}
\affiliation{University of Potsdam, Institute of Physics and Astronomy, Karl--Liebknecht--Str. 24--25, 14476 Potsdam, Germany}
\affiliation{Department of Physics and Astronomy, University of Exeter, Stocker Road, Exeter EX4 4QL, UK}

\begin{abstract}
Magnetization dynamics is commonly described by the stochastic Landau–Lifshitz–Gilbert (LLG) equation. 
On picosecond timescales, inertial and open-system extensions of the LLG equation are necessary to interpret recent experiments.
We show analytically and numerically that the standard LLG equation exhibits strictly positive entropy production rates, while inertial and open-system LLG dynamics display temporarily negative entropy production rates indicating non-Markovianity.
Here we quantify the degree of non-Markovianity using established measures.
Our numerical calculations show that the open-system LLG equation consistently exhibits the highest magnitude of non-Markovianity for different initial conditions and magnetic field orientations. 
\end{abstract}
\maketitle

\section{Introduction}
\label{sec:intro}
Translational Brownian motion is one of the core concepts to describe fluids~\cite{Tothov_2011}, gases~\cite{Schatz_1978, Green_1951}, and a wide range of processes in statistical physics~\cite{Satija_2019}, e.g., the random motion of micro-sized particles~\cite{Perez_Guerrero_2024, Verweij_2023} and molecules~\cite{Serag_2017} dispersed in aqueous solutions.
Generally speaking, translational Brownian motion is modeled via a generalized Langevin equation, which includes memory effects, i.e., time-dependent friction, and colored noise fluctuations.
These memory effects and fluctuations are linked by the fluctuation-dissipation theorem and lead to non-Markovian dynamics~\cite{Loos2025}.
In many physical situations, the assumption of linear (time-independent) friction is sufficient, and the dynamics becomes instantaneous, i.e., Markovian. 
That implies that the immediate next probability distribution of the system $p(t+\delta t)$ only depends on the current probability distribution $p(t)$, but not on any earlier probability distributions $p(t- a\delta t)$ ($a\in \mathrm{N}$)~\cite{livi2017nonequilibrium}, $t$ denotes time and $\delta t$ an infinitesimal timestep.
This Markovian condition is satisfied if the bath relaxes significantly faster than the system~\cite{Breuer_2007}, $\tau_{\mathrm{B}} \ll \tau_{\mathrm{S}}$, where $\tau_{\mathrm{B}}$ and $\tau_{\mathrm{S}}$ are the bath and system relaxation times, respectively. 
If the system (strongly) interacts with a structured or small thermal heat bath~\cite{Wenderoth_2021}, then time-scale separation is broken~\cite{Breuer_2007, Volkov_1996, Rivas_2014}, $\tau_{\mathrm{B}} \sim \tau_{\mathrm{S}}$, and energy from the bath may flow back into the system.
Therefore, the immediate next probability distribution of the system $p(t+\delta t)$ does not only depend on the current probability distribution $p(t)$, but might depend on any combination of previous probability distributions $p(t - a\delta t)$~\cite{livi2017nonequilibrium}, i.e., the dynamics is non-Markovian.

Here we consider a different kind of Brownian motion, namely rotational Brownian motion. In this case, the motion is constrained to the surface of a sphere rather than occurring in unconstrained Euclidean space as in translational Brownian motion~\cite{Miyazaki_1998, Garca_Palacios_1998, Hoefling_2025}.
Such rotational motion occurs in nature, e.g. in the magnetization dynamics in ferromagnets. 

In contrast to the generalized Langevin equation for translational Brownian motion, the rotational generalized Langevin equation~\cite{Miyazaki_1998} is non-linear~\cite{Zwanzig_1973} and the noise enters in a multiplicative way, making its analytical treatment more challenging~\cite{Tranchida_2016, Ma_2011, Tranchida_2016_2}.
One special case of it is the Landau-Lifshitz-Gilbert (LLG) equation~\cite{Landau_1965,gilbert2004phenomenological}.
It is the most common stochastic magnetization dynamics equation, which is used to describe damped precessional motion in an effective magnetic field~\cite{Lakshmanan_2011}.
The LLG equation has been extended by an inertial term~\cite{Ciornei_2011, Olive_2012, Faehnle_2011, Thonig_2017, mondal2018generalisation, Mondal_2017, Thibaudeau_2021, Quarenta_2024} to the so-called inertial LLG (iLLG) equation~\cite{Mondal_2023}.
The inertial term leads to an additional oscillation, i.e., nutation, that has been observed experimentally on ultrafast timescales~\cite{neeraj2021inertial, De_2025}.
Recent experiments on crystalline cobalt thin films have measured not one but several additional oscillations in magnetization dynamics~\cite{unikandanunni2022inertial, Hartmann_2025}.
All these oscillations
are captured by a \textit{generalized}~\footnote{The equation is referred to as general in the sense of the generalized Langevin equation for translational Brownian motion.} magnetization dynamics equation, which has been derived from an open-quantum system approach~\cite{Anders2022quantum, Hartmann_2025}.
Here, we refer to it as the open-system Landau-Lifshitz-Gilbert equation (os-LLG)~\cite{Weber_2025}. 

Thermodynamic entropy production rates (EPR) are a tool to investigate non-Markovianity in ultrafast magnetization dynamics~\cite{Caprini_2024, tietjen2025ultrafast,bandopadhyay2015stochastic}.
In quantum systems, the entropy production rate $\dot{\Sigma}(t)$ can be negative due to quantum fluctuations, entanglement and correlations between system and bath, and backflow of energy~\cite{esposito2010entropy,marcantoni2017entropy,latune2020negative,kutvonen2015entropy}.
In classical systems, negative entropy production rates indicate a non-Markovian evolution if the system is undriven and local detailed balance holds~\cite{strasberg2019non}. 
The entropy production rate can be defined via the relative entropy $D$ (often also referred to as Kullback-Leibler divergence)
\begin{equation}\label{eq:rel-EP}
    D[p_\alpha(t)||\pi_\beta] =  \sum_\alpha p_\alpha(t) \ln\frac{p_\alpha(t)}{\pi_\beta},
\end{equation}
where $k_{\mathrm{B}}$ is the Boltzmann constant.
It is a measure of the distinguishability between two probability distributions $p_\alpha$ and here the thermal equilibrium state $\pi_\beta$, which is the fixed point of the dynamics and necessary for the calculation of the entropy production rate~\cite{strasberg2019non}.
It is defined as $\pi_\beta = e^{-\beta E_\alpha}/\mathcal{Z}$, where $\beta$ and $\mathcal{Z} = \sum_\alpha e^{-\beta E_\alpha}$ are the inverse temperature and the partition function, respectively~\cite{strasberg2019non}. 
The energy of the system in state $\alpha$ is denoted by $E_\alpha$.
The negative time derivative of the relative entropy defines then the entropy production rate $\dot{\Sigma}$~\cite{altaner2017nonequilibrium,strasberg2019non}
\begin{equation}\label{eq:sigma=KL}
    \dot{\Sigma}(t) = - k_{\mathrm{B}}\,\partial_t D[p_\alpha(t)||\pi_\beta].
\end{equation}

The following theorem linking entropy production rates and non-Markovianity was proven by Strasberg and Esposito~\cite{strasberg2019non}: If a classical system is undriven, local detailed balance holds, and the total system is initialized within the set of conditional steady states, then the occurrence of negative entropy production rates indicate that the system undergoes a non-Markovian evolution.
Note that positive entropy production rates do not necessarily imply purely Markovian dynamics.

In this work, we determine the entropy production rates~\cite{esposito2010entropy,strasberg2019non,Esposito_2012,Landi_2021} of the LLG, iLLG, and os-LLG equations and use them to detect and quantify the non-Markovianity of each magnetization dynamics.
We analytically and numerically show that the LLG equation only leads to positive entropy production rates~\cite{breuer2016colloquium}.
In contrast, we show that the entropy production rates of the iLLG and os-LLG equations can be negative, signaling a non-Markovian evolution~\cite{strasberg2019non}.
By employing two different non-Markovianity measures, ``negative EPR window''~\cite{Laine_2010} and ``relative EPR window'', we quantitative compare the differences of the LLG, iLLG and os-LLG equation dynamics.

We consistently find that the magnitude of non-Markovianity is largest for the os-LLG equation, which indicates that it is the most non-Markovian of the magnetization dynamics compared here.
While the LLG equation exhibits zero non-Markovianity regardless of the initial conditions, the iLLG equation exhibits non-Markovianity for a canted orientation between the external field and initial magnetization.
This indicates a dependence of the non-Markovianity on the initial conditions.

\section{Ultrafast magnetization dynamics}
\label{sec:mag_dyn}
Stochastic magnetization dynamics on ultrafast timescales is commonly described by the LLG equation
\begin{equation}\label{eq:classic-LLG}
    \dot{\mathbf{m}} = - \gamma \mathbf{m} \times \left( \mathbf{B}_{\mathrm{eff}} + \mathbf{b}_{\mathrm{th}}(t)  + \alpha \dot{\mathbf{m}} \right)
\end{equation}
where $\mathbf{m}$ is the unit-free, normalized magnetization vector, which precesses because the torque interaction with the effective magnetic field $\mathbf{B}_{\mathrm{eff}}$, scaled by the electronic gyromagnetic ratio $\gamma$.
This field can consist of any combination of a constant external bias field, a possibly time-dependent external field, and any anisotropy field arising from the material and magnetization itself.
The damping in the third term is scaled by the Gilbert damping parameter $\alpha$. 
The thermal noise is denoted by $\mathbf{b}_{\mathrm{th}}$ and is Gaussian white noise with 
\begin{align*}
    \langle \mathbf{b}_{\mathrm{th}}(t) \rangle &= 0 \\ 
    \langle b_{\mathrm{th},i}(t)\, b_{\mathrm{th},j}(t')\rangle &= \eta\,\delta_{ij}\,\delta(t-t') \,,
\end{align*}
where the correlation factor is $\eta= \alpha k_BT$.

The stochastic LLG Eq.~\eqref{eq:classic-LLG} can be extended by an inertial term~\cite{Ciornei_2011, Olive_2012, Faehnle_2011, Thonig_2017, mondal2018generalisation, Mondal_2017, Thibaudeau_2021, Quarenta_2024}
which is scaled by the angular momentum relaxation time $\tau$ and describes an additional nutation motion
\begin{equation} \label{eq:inertial-LLG}
    \dot{\mathbf{m}} = - \gamma \mathbf{m} \times \left( \mathbf{B}_{\mathrm{eff}} + \mathbf{b}_{\mathrm{th}}(t)  + \alpha \dot{\mathbf{m}} + \alpha\tau \ddot{\mathbf{m}} \right)\,.
\end{equation}

Both of these stochastic magnetization Eqs.~\eqref{eq:classic-LLG} and~\eqref{eq:inertial-LLG} are contained in the os-LLG equation~\cite{Anders2022quantum}, which has been shown to successfully describe higher-order oscillations in ultrafast dynamics~\cite{Hartmann_2025} and to produce the correct equilibrium magnetization~\cite{Weber_2025}.
The os-LLG equation arising from an isotropic coupling of the magnetic system with a thermal bath is given by
\begin{equation}
    \dot{\mathbf{m}} = - \gamma \mathbf{m}\times\left( \mathbf{B}_{\mathrm{eff}} + \tilde{\mathbf{b}}_{\mathrm{th}}(t) + \int
    _{-\infty}^{t}\mathrm{d}t'\, \mathcal{K}(t-t')\mathbf{m}(t') \right),
    \label{eq:open-system LLG}
\end{equation}
where $\tilde{\mathbf{b}}_{\mathrm{th}}$ describes again the thermal fluctuations of the bath.
However, these fluctuations are now colored (for better clarity, denoted by a tilde) and $\mathcal{K}(t-t')$ encodes the memory of the magnetization dynamics. 
Importantly, here, the power spectrum of the fluctuations $P(\nu)$ and the memory kernel (dissipation) are linked by the fluctuation-dissipation theorem, i.e.~$P(\nu) = 2k_{\mathrm{B}}T\,\mathrm{Im[\mathcal{K}(\nu)]}$, with $\mathcal{K}(\nu)$ being the Fourier transform of the memory kernel $\mathcal{K}(t-t)$.

We note that all three LLG equations preserve the magnitude of magnetization based on their structure.
Without loss of generality, we choose $|\mathbf{m}| = 1$ for all equations.

\section{Analytical Bounds for Entropy Production Rates}
\label{sec:analytics}
We now analyze \textit{contractivity} ~\cite{strasberg2019non} for the entropy production rates of the different magnetization dynamics analytically.
Contractivity states that two probability distributions will monotonically come closer in time for all times, implying
\begin{equation}
    \dot{\Sigma}(t) \geq 0,
    \label{eq:contractivity}
\end{equation} with $\dot{\Sigma}(t) = 0$ in thermal equilibrium.
Contractivity is therefore directly linked to the theorem formulated by Strasberg and Esposito~\cite{strasberg2019non}, when contractivity is broken the dynamics is non-Markovian.

For the LLG Eq.~\eqref{eq:classic-LLG}, we prove that the entropy production rate is always positive, i.e., the magnetization dynamics is contractive.
To show this, consider the definition of the Fokker-Planck equation for the stochastic LLG equation~\cite{Garca_Palacios_1998}
\begin{align}
    \frac{\partial p(\mathbf{m},t)}{\partial t} =& -\nabla_{S^2} \cdot\left(\mathbf{A}(\mathbf{m}) p(\mathbf{m},t) \right) \nonumber\\
    &+ \frac{1}{2}\nabla_{S^2}\cdot\left(\mathcal{D}(\mathbf{m})\nabla_{S^2}\,p(\mathbf{m},t) \right),
    \label{eq:FPE_short}
\end{align}
where $\mathbf{A}(\mathbf{m}) = - \gamma \mathbf{m}\times\mathbf{B}_{\mathrm{eff}} - \gamma\alpha\mathbf{m}\times(\mathbf{m}\times\mathbf{B}_{\mathrm{eff}})$ is the drift vector field of the deterministic part of the dynamics and $\mathcal{D}(\mathbf{m}) = 2\alpha k_{\mathrm{B}}T (\mathbb{1}-\mathbf{m}\mathbf{m}^\mathrm{T})/\gamma$ is the diffusion tensor (for details see App.~\ref{app:FPE_LLG}). 
The operator $\nabla_{S^2}$ denotes differentiation with respect to the surface of the unit sphere $S^2$, and $ p(\mathbf{m},t)$ is the probability distribution for magnetization $\mathbf{m}$ at time $t$.

The entropy production rate Eq.~\eqref{eq:sigma=KL} in its continuous version with integration over the full phase-space $S^2$ reads
\begin{equation}
    \frac{\dot{\Sigma}(t)}{k_{\text{B}}} = -\frac{\partial D(t)}{\partial t} = -\int_{S^2} \ln \left(\frac{p(\mathbf{m},t)}{\pi_\beta(\mathbf{m})}\right)\frac{\partial p(\mathbf{m},t)}{\partial t}\mathrm{d}\Omega.
    \label{eq:D_KL_derivative_short}
\end{equation}
Plugging the FPE of the LLG equation Eq.~\eqref{eq:FPE_short} into Eq.~\eqref{eq:D_KL_derivative_short} yields a quadratic form, such that $\dot{\Sigma}(t) \geq 0$ for all times $t$.
Details of this proof are given in App.~\ref{app:markov-LLG}. 

\begin{figure*}
    \centering
    \includegraphics[width=0.98\textwidth]{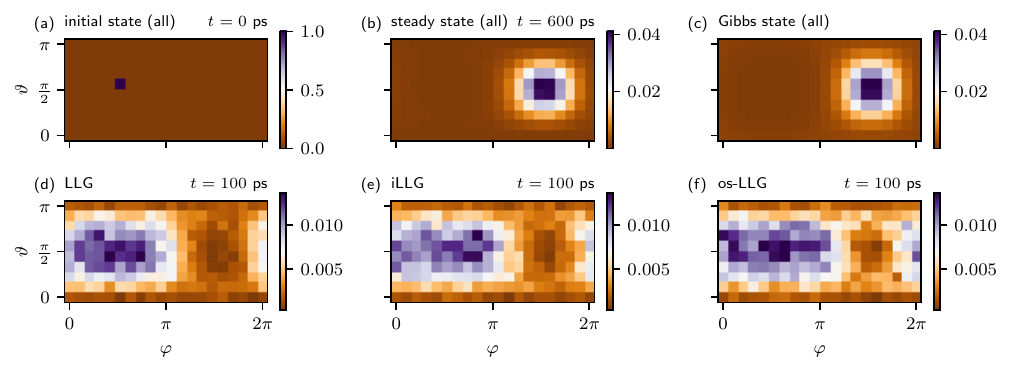}
    \caption{\textbf{Probability distributions of the magnetization vector in spherical coordinates for LLG, iLLG, and os-LLG:} (a) For all three equations, the magnetization is initialized in the parallel configuration.
    (b) The steady state distribution is obtained by time-averaging the distribution once the system has equilibrated, here exemplarily shown for the LLG equation.
    (c) Analytically calculated Gibbs state $\pi_\beta(\mathbf{m})$, which matches the numerically obtained steady-state of each of the magnetization equations. 
    (d), (e), (f) Shows the magnetization distributions after $t = 100$~ps for the LLG, iLLG, and os-LLG equations, respectively, evidencing differences in the temporal evolution. 
    All six panels show the distribution of the magnetization in spherical coordinates $(\varphi,\vartheta)$ for a fixed magnetization magnitude of $\abs{\mathbf{m}} = 1.0$.
    }
    \label{fig:distributions}
\end{figure*}

The iLLG Eq.~\eqref{eq:inertial-LLG}, on the other hand, does not always lead to a contractive dynamics as defined in Eq.~\eqref{eq:contractivity}.
This is because the full process is no longer described by the (marginal) probability density $p(\mathbf{m},t)$ but by a probability density $P(\mathbf{m},\mathbf{v},t)$, which arises from the second-order time derivative in Eq.~\eqref{eq:inertial-LLG}.
By integrating over the ``velocity'' variable $\mathbf{v}$, using the Mori-Zwanzig formalism~\cite{Vrugt_2020,Breuer_2007} as done in App.~\ref{app:iLLG_osLLG_KL}, the complete parameter space is reduced and information about the dynamics is lost. 
Therefore, the evolution of the marginal probability density $p(\mathbf{m},t)$ is no longer time-local, and memory effects are introduced.
Hence, the iLLG Eq.~\eqref{eq:inertial-LLG} describes a non-Markovian evolution of the magnetization vector $\mathbf{m}$ and, in general, $\dot{\Sigma}(t) \geq 0$ no longer holds.

Similar to the iLLG case, for the os-LLG Eq.~\eqref{eq:open-system LLG} the full probability density is a higher-dimensional function $P(\mathbf{m},\mathbf{v},\mathbf{w},t)$. 
Thus, a similar argument can be constructed to show that the os-LLG Eq.~\eqref{eq:open-system LLG} leads to a non-Markovian evolution for the magnetization $\mathbf{m}$ alone, see App.~\ref{app:iLLG_osLLG_KL}.
Having established that both iLLG and os-LLG can break contractivity, we will quantify their degree of non-Markovianity below.

\section{Non-Markovianity measures}
\label{sec:entropy_rates}
To do so we need non-Markovianity measures. 
Here we choose to quantify the degree of non-Markovianity using the entropy production rate.
We employ the following, previously characterized and used measure~\cite{Laine_2010}, which we refer to as ``negative EPR window''
\begin{equation}
    \mathcal{N}_{\mathrm{A}} = -\int_{\dot\Sigma(t) <0}\dot{\Sigma}(t)\,\mathrm{d}t \,,
    \label{eq:nonMarkov_measure_area}
\end{equation} 
where the time integration is over all intervals in which $\dot{\Sigma}(t) < 0$. 
Consequently, for a Markovian process, where $\dot\Sigma(t) \geq 0$ for all times, we have $\mathcal{N}_{\mathrm{A}} = 0$.
This measure also reflects the absolute strength of the relative entropy, which can be of interest but makes comparison between different setups difficult.
Therefore, we also introduce the normalized version as a non-Markovian measure, here called ``relative EPR window'', 
\begin{equation}
    \overline{\mathcal{N}}_{\mathrm{A}} = -\frac{\int_{\dot\Sigma(t) <0}\dot{\Sigma}(t)\,\mathrm{d}t}{\int|\dot{\Sigma}(t)|\,\mathrm{d}t} \,,
    \label{eq:nonMarkov_measure_area_norm}
\end{equation}
where we have normalized by the total time integral of the absolute value of the entropy production rate, which is equal to the absolute area of the entire entropy production rate curve.
This leads to a measure that is independent of the absolute scale of the relative entropy.
By definition, $\overline{\mathcal{N}}_{\mathrm{A}}$ takes values between $[0,1]$ with $\overline{\mathcal{N}}_{\mathrm{A}} > 0$ indicating non-Markovian dynamics.

There are other Markovianity measures~\cite{Addis_2014}, which we leave as tools for future work. 
Two examples, counting timesteps with negative entropy production rate and counting the number of sign switches of the entropy production rate, are briefly discussed in App.~\ref{app:measures}.

\section{Entropy Production Rates and non-Markovianity for different Magnetization Dynamics}
\label{sec:num_results}

In what follows, we numerically investigate the relative entropy and the entropy production rate as defined in Eqs.~\eqref{eq:rel-EP} and~\eqref{eq:sigma=KL} and quantify non-Markovianity via the negative and relative EPR windows. 
We simulate $N_{\mathrm{traj}} = 50000$ trajectories of the LLG Eq.~\eqref{eq:classic-LLG}, iLLG Eq.~\eqref{eq:inertial-LLG} and os-LLG Eq.~\eqref{eq:open-system LLG} equations with realistic parameter sets for cobalt thin films measured in ultrafast magnetization dynamics experiments~\cite{unikandanunni2022inertial, Hartmann_2025}.

To specify the dynamics of the os-LLG equation, we need to choose a spectral density $I(\nu)$, which details the coupling of the system to the bath and relates to the memory kernel via $\mathrm{Im}[\mathcal{K}(\nu)] = I(\nu)$.
As in previous works, we use a Lorentzian spectral density~\cite{Hartmann_2025,Nemati_2022},~$I(\nu) = \alp\Gamma\nu/((\nu_0^2-\nu^2)^2+\nu^2\Gamma^2)$,
where the $\nu_0$ denotes the central frequency and $\Gamma$ the width of the Lorentzian. 
The amplitude $\alp$ is proportional to the strength of the system-bath interaction and the dissipation. 
The Lorentzian spectral density describes, e.g., the coupling of the magnetization to modes of the phonon bath.
The Lorentzian parameters are fixed as $\alp = 242\,\mathrm{THz}^3$, $\nu_0 = 4.2$~THz and $\Gamma = 0.2$~THz and relate to the parameters of the LLG and iLLG via $\alpha = \alp\Gamma/\nu_0^4$ and $\tau = (\nu_0^2-\Gamma^2)/(2\pi\nu_0^2\Gamma)$~\cite{Hartmann_2025}. 
Hence, we have a damping parameter of $\alpha =0.15$ and an inertial parameter of $\tau = 120$~fs.

By the theorem of Ref.~\cite{strasberg2019non} (see Sec.~\ref{sec:intro}), the initial conditions of the simulations need to be in the class of conditional steady states in order to conclude that negative entropy production rates imply non-Markovian dynamics.
We therefore selected our initial conditions accordingly for the numerical results discussed below.

\begin{figure}
    \centering
    \includegraphics[width=0.48\textwidth]{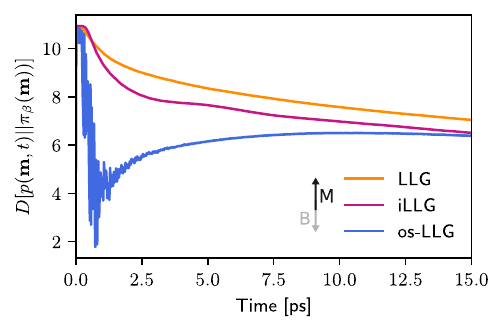}
    \caption{\textbf{Relative entropy of the LLG, iLLG, and os-LLG equations} initialized in the parallel configuration as a function of time.
    The damping parameter is always $\alpha=0.15$ and the inertial parameter is $\tau = 120$~fs for the simulations of iLLG and os-LLG (with Lorentzian parameters fixed as $\nu_0 = 4.2$~THz, $\Gamma = 0.2$~THz and $\alp = 242\,\mathrm{THz}^3$, see Ref.~\cite{Hartmann_2025}). For all $N_{\mathrm{traj}} = 20000$ trajectories the temperature is $T_{\mathrm{sim}} = 0.1$.
    }
    \label{fig:EP_example}
\end{figure}

In the following we will characterize the dynamics starting from two initial configurations, which capture two different geometries and field strengths: 
\begin{itemize}
    \item parallel: $\mathbf{m}_0 = (0,1.0,0)$, $\mathbf{B}_{\mathrm{eff}} = -0.35\mathrm{T}\,\hat{\mathbf{y}}$;
    \item canted: $\mathbf{m}_0 = (\sqrt{0.5},-\sqrt{0.5},0)$, $\mathbf{B}_{\mathrm{eff}} = -1.5\mathrm{T}\,\hat{\mathbf{y}}$.
\end{itemize}
The first configuration starts with anti-parallel alignment between $\mathbf{B}_{\mathrm{eff}}$ and $\mathbf{m}_0$, while the canted configuration has a $45^{\circ}$ angle between field and magnetization, as well as a stronger $\mathbf{B}_{\mathrm{eff}}$.
In all simulations, we choose a unitless temperature of $T_{\mathrm{sim}}=0.1$.

We obtain the probability distribution of the magnetization $p(\mathbf{m},t)$ by transformation of the normalized magnetization vector into spherical coordinates $\mathbf{m} = (r=\abs{\mathbf{m}},\varphi,\vartheta)$ with $\varphi \in [0,2\pi)$ and $\vartheta \in[0,\pi]$, and $\abs{\mathbf{m}}=1$ remains unchanged by the dynamics. 
Next, we discretise the phase-space $(\varphi,\vartheta)$ by $\varphi/30$ and $\vartheta /15$ to obtain $450$ bins (see App.~\ref{app:coarse_graining}) and count at each time step $t_n$ how often $\mathbf{m}$ lies within a certain bin. 
The total number of trajectories normalizes the counts to get a properly normalized probability distribution $p(\mathbf{m},t)$, as shown in Fig.~\ref{fig:distributions}.

We start with a fixed initial magnetization (Fig.~\ref{fig:distributions}~(a)) and let the system evolve until it reaches equilibrium.
Regardless of the equation that governs the dynamics, Fig.~\ref{fig:distributions}(b) and (c) illustrate that the numerically obtained steady state (after $600$~ps) is the Gibbs state, i.e., the expected magnetization configuration in thermal equilibrium (for further details see App.~\ref{app:gibbs}).
To calculate the relative entropy, we therefore use the Gibbs state $\pi_\beta(\mathbf{m})$, which is known analytically and whose numerically binned version we can obtain directly.
The intermediate magnetization configurations (after $100$~ps) for the LLG, iLLG and os-LLG equations are shown in Fig.~\ref{fig:distributions}~(d), (e) and (f), respectively.
They differ and therefore confirm that the three equations describe different dynamics.
Knowing $p(\mathbf{m},t)$ at each time step $t_n$, we calculate the relative entropy $D(p(\mathbf{m},t)||\pi_\beta(\mathbf{m}))$ and the entropy production rate $\dot{\Sigma}(t)$ for all three simulated equations of motion.

The calculated relative entropy for the parallel configuration during the first $15$~ps is shown in Fig.~\ref{fig:EP_example}. 
As expected from the analytical discussion in Sec.~\ref{sec:mag_dyn}, we observe very different behavior of the relative entropy during the first $\approx 15$~ps of the three different magnetization equations.
While the relative entropy for the LLG equation decreases monotonously, the relative entropy for the iLLG equation shows small variations, and the relative entropy for the os-LLG equations is clearly non-monotonous, where the high-frequency oscillations stem from the memory kernel in Eq.~\eqref{eq:open-system LLG}.
All three cases become strictly monotonously declining for larger times, $t > 15$~ps, (see Fig.~\ref{fig:DKL_binning} in App.~\ref{app:coarse_graining}), indicating that the non-Markovian dynamics is present only on short, ``ultrafast'' timescales.
As we show in App.~\ref{app:coarse_graining}, the overall magnitude of the numerically obtained relative entropy depends on the binning we choose. 
Therefore, what matters here are the relative differences between the three LLG-type dynamics Eqs.~\eqref{eq:classic-LLG},~\eqref{eq:inertial-LLG}~and~\eqref{eq:open-system LLG}.

\begin{figure}
    \centering
    \includegraphics[width=0.48\textwidth]{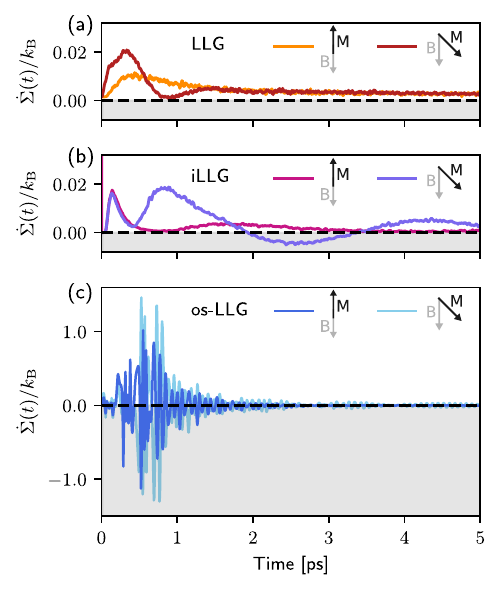}
    \caption{\textbf{Entropy production rate of the LLG, iLLG, and os-LLG equation.} 
    (a) Entropy production rates of the LLG equation for parallel and canted configurations stay positive for all times, while (b) for the iLLG equation, the canted configuration leads to temporarily negative entropy production rates.
    (c) The entropy production rates of the os-LLG equation oscillate quickly between positive and negative for both, parallel and canted configurations.
    The grey region highlights violations of contractivity.
    Traces shown are after taking a moving average over five data points that suppresses the fluctuations of the limited statistics (50~000 trajectories). The remaining parameters correspond to those in Fig.~\ref{fig:EP_example}.
    }
    \label{fig:EPR_comparison}
\end{figure}

In Fig.~\ref{fig:EPR_comparison}, we compare the calculated entropy production rates $\dot{\Sigma}(t)$ for all three magnetization dynamics equations and both parameter configurations.
Since the distribution $p(\mathbf{m},t)$ is made up of a finite sample of $N_{\mathrm{traj}} = 50000$ trajectories, the numerically obtained entropy production rate shows visible fluctuations. 
To reduce the noise, we apply a moving time average on the entropy production rate over five data points.
While some sampling noise remains, one can see that the entropy production rate of the LLG equation stays strictly positive for both configurations (staying above the black dashed line in Fig.~\ref{fig:EPR_comparison}(a)), while the entropy production rate of the iLLG temporarily becomes negative for the canted configuration, but stays strictly positive for the parallel one (see Fig.~\ref{fig:EPR_comparison}(b)).
This is mainly attributed to the torque in the magnetic field that the magnetization vector initially experiences, which is larger for the canted configuration than the parallel one~\cite{Ciornei_2011,neeraj2021inertial}.
For the os-LLG Eq.~\eqref{eq:open-system LLG}, the entropy production rate becomes temporarily negative in both configurations and changes sign frequently (Fig.~\ref{fig:EPR_comparison}(c)).
The magnitude of the entropy production rate is larger than that of the LLG and iLLG equations, which is due to the dynamics of the total system-bath setup. 

\begin{figure}
    \centering
    \includegraphics[width=0.48\textwidth]{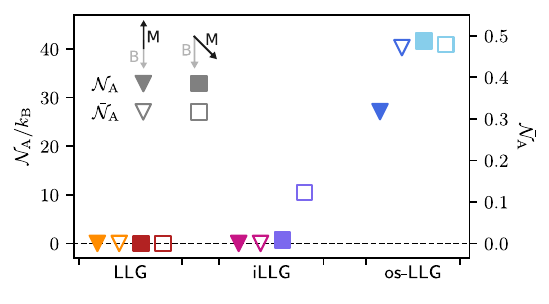}
    \caption{\textbf{Magnitude of non-Markovianity.}  For the three different magnetization dynamics equations and the two different configurations (triangles stand for parallel and squares for canted configuration), we calculate the non-Markovianity measures $\mathcal{N}_A$~Eq.~\eqref{eq:nonMarkov_measure_area} (filled symbols) and $\overline{ \mathcal{N}}_A$~Eq.~\eqref{eq:nonMarkov_measure_area_norm} (empty symbols). 
    For both measures, they are zero for both configurations in the LLG case. In the iLLG case, they are zero for the parallel configuration but non-zero for the canted configuration. In the os-LLG case, both non-Markovianity measures are non-zero for both configurations and significantly larger than in the iLLG case.
    The difference between the configurations is larger in the iLLG case, but smaller in the os-LLG case. 
    The measures $\mathcal{N}_{\mathrm{A}}$ and $\overline{\mathcal{N}}_{\mathrm{A}}$ are evaluated over the first $12$~ps of the dynamics for the trajectories shown in Fig.~\ref{fig:EPR_comparison}.}
    \label{fig:EP_measure_area}
\end{figure}

To further quantify the differences in the entropy production rate, we employ the discrete versions of Eq.~\eqref{eq:nonMarkov_measure_area} and Eq.~\eqref{eq:nonMarkov_measure_area_norm} to calculate the negative EPR window $\mathcal{N}_{\mathrm{A}}$ and the relative EPR window $\overline{\mathcal{N}}_{\mathrm{A}}$ from the entropy production rates presented in Fig.~\ref{fig:EPR_comparison}.
The resulting values are plotted in Fig.~\ref{fig:EP_measure_area}.
For the parallel configuration, we find that both EPR windows are zero for the LLG and iLLG equations, while it is non-zero for the os-LLG equation.
For the canted configuration, only the LLG equation has a zero measure, while iLLG and os-LLG have finite values.
The negative EPR window of the os-LLG equation has a value of $\mathcal{N}_{\mathrm{A}}\sim 30$, which is one order of magnitude larger than for the iLLG equation $\mathcal{N}_{\mathrm{A}}\sim 2$.

We remark that we have introduced a small threshold of $\epsilon = -0.001$, such that we only counted the regions in which $\dot{\Sigma}(t) < \epsilon$ holds. This is again to reduce fluctuations stemming from the finite number of simulated trajectories.

The relative EPR window $\overline{\mathcal{N}}_{\mathrm{A}}$ additionally reveals that the non-Markovianity in the iLLG is significantly stronger for the canted configuration than for the parallel one.
The results for $\overline{\mathcal{N}}_{\mathrm{A}}$ also show that the non-Markovianity of the os-LLG does not differ as much between parallel and canted configurations as it does for the iLLG case.
This underlines the non-Markovianity of the os-LLG beyond the inertial term, because the stronger excitation of the inertial term in the canted configuration does not lead to significant changes in the non-Markovianity.

The numerical results exemplify and complete the analytical results discussed in Sec.~\ref{sec:mag_dyn} and in App.~\ref{app:markov-LLG} and~\ref{app:iLLG_osLLG_KL}.
If the magnetic system of interest is assumed to evolve via the standard LLG Eq.~\eqref{eq:classic-LLG}, it must show a strictly positive entropy production rate. 
In contrast, if the magnetic system of interest exhibits a negative entropy production rate, the system undergoes non-Markovian evolution, and memory effects must be accounted for, e.g., using the os-LLG equation.

\section{Discussion and Conclusion}
\label{sec:discussion_conclusion}
In summary, our manuscript uses the thermodynamic concept of negative entropy production rates to detect and characterize non-Markovian (memory-dependent) magnetization dynamics on ultrashort timescales~\cite{Walowski_2016,Rossi_2002} for three different equations of motion, i.e., LLG Eq.~\eqref{eq:classic-LLG}, iLLG Eq.~\eqref{eq:inertial-LLG}, and os-LLG Eq.~\eqref{eq:open-system LLG}.
Analytical arguments showed that the LLG equation always has a positive entropy production rate, whereas the entropy production rates of the iLLG and os-LLG equations may temporarily become negative, depending on the initial conditions and magnetic field strengths.

The numerically calculated relative entropy of all three equations directly showed that they are different on short timescales (up to approximately $10$~ps), while they converge in the long time limit.
Using two non-Markovianity measures, which are based on the negativity of the entropy production rates, confirmed that the os-LLG Eq.~\eqref{eq:open-system LLG} shows the largest magnitude of non-Markovianity. 
This evidences that by including a memory kernel in the os-LLG equation, the magnetization dynamics can become non-Markovian~\cite{Anders2022quantum,Hartmann_2025,Verstraten_2023}. 
Similarly, the iLLG Eq.~\eqref{eq:inertial-LLG} dynamics can lead to negative entropy production rates. 
As expected from the analytical calculations, we numerically found that the LLG Eq.~\eqref{eq:classic-LLG} leads to purely positive entropy production rates and a vanishing non-Markovianity measure.

We note again that the details of the relative entropy and entropy production rate depend on the material parameters, the initial condition, the temperature and the statistics of the thermostat~\cite{Weber_2025}, and the different internal and external magnetic fields. 
However, whenever the magnetization dynamics is found to yield a negative entropy production rate (given the dynamics is undriven, detailed balance holds, and the fixed point of the dynamics is known), the correct theoretical framework cannot be the LLG equation.

Furthermore, we want to remark that we used multiplicative noise as derived in Ref.~\cite{Miyazaki_1998,Anders2022quantum,Suhl_1998} in our simulations.
Alternatively, one could include additive noise~\cite{Ma2012longitudinal}, which introduces longitudinal fluctuations along the magnetization vector and enables the path-integral formalism~\cite{tietjen2025ultrafast}.
This adds another dynamic variable, and it might change the observed entropy production rates.

While non-Markovianity in magnetic systems has been studied at least to some extent theoretically~\cite{Miyazaki_1998,Atxitia_2009,Verstraten_2023,ReyesOsorio_2025,popovic2018entropy,Mchugh_2018}, only recently has this been connected to an experiment on ultrafast timescales~\cite{Hartmann_2025}. 
Further ultrafast experiments in (magnetic) solid state materials may probe and detect non-Markovian evolution of magnetic or non-magnetic degrees of freedom in the future. 
One promising direction to experimentally access the entropy production is to use hot-electron bolometers~\cite{Karimi_2024}.
This could reveal new physics as well as lead to new technological applications. 

\section*{Acknowledgements}
We would like to thank Yulong Qiao, Karen Hovhannisyan, and Fried-Conrad Weber for fruitful discussions.
FH and JA acknowledge support from the DFG, grants
513075417 and 384846402. JA acknowledges support
from EPSRC grant EP/R045577/1, and the Royal Society.
FT and RMG acknowledge support from the Swedish Research Council (VR starting Grant No. 2022-03350), the Olle Engkvist Foundation (Grant No. 229-0443), the Royal Physiographic Society in Lund (Horisont), the Knut and Alice Wallenberg Foundation (Grant No. 2023.0087), and Chalmers University of Technology, via the department of physics and the Areas of Advance Nano and Materials Science. 

\section*{Author contributions}
FH and FT discussed and developed the initial idea. FH did the analytical and numerical calculations of the entropy production rates. FT developed code to numerically solve the iLLG equation.
FT (lead) and FH (supportive) wrote a first draft of the manuscript. 
All authors finalized the paper. RMG and JA supervised the project.
\section*{Competing interests}
The authors declare no competing interests.
\section*{Data availability}
The code and the generated numerical data is available upon reasonable request from the corresponding authors.

\bibliography{main}

\newpage
\clearpage

\onecolumngrid

\appendix

\section*{Appendix}
The first part of this Appendix (App.~\ref{app:cont_LLG} and App.~\ref{app:iLLG_osLLG_KL}) gives the full derivation of the theorem of contractivity as discussed in Sec.~\ref{sec:analytics} in the main text. 
It states that the undriven LLG Eq.~\eqref{eq:classic-LLG} always leads to a contractive process Eq.~\eqref{eq:contractivity}, such that the entropy production rate obeys $\dot{\Sigma}(t)\geq 0$ for all times $t$.
In contrast, contractivity generally no longer applies to the iLLG Eq.~\eqref{eq:inertial-LLG} and os-LLG Eq.~\eqref{eq:open-system LLG}, so that $\dot{\Sigma}(t)$ can temporarily become negative.
The general argument is as follows:

Stochastic processes are fully characterized by their probability density. If the \textit{full} probability density governing the dynamics is taken into account, i.e., the description is closed and includes all relevant degrees of freedom, then the entropy production rate is always non-negative and the dynamics is contractive Eq.~\eqref{eq:contractivity} towards the equilibrium probability density, i.e., on this level the dynamics is completely Markovian~\cite{Kupferman_2004,Siegle_2010,livi2017nonequilibrium}.
However, if the \textit{full} probability density is unknown or is experimentally inaccessible, and only marginal distributions of a reduced set of variables are considered, 
the resulting effective dynamics does not need to be contractive. 
In this case, entropy production rates obtained from marginal distributions alone can become negative, which reflects the influence of unresolved degrees of freedom~\cite{strasberg2019non}.
This is a general statement about stochastic processes, which is independent of the underlying physical system. 

We now apply this argument to (ultrafast) magnetization dynamics.
For the LLG equation the \textit{full} probability density $p(\mathbf{m},t)$ is given by the magnetization $\mathbf{m}$ at any time $t$ alone. No further variables guide the evolution of the magnetization vector and $p(\mathbf{m},t)$ contains the full information. 
This will lead to a contractive process at any given time, as we show in App.~\ref{app:cont_LLG}.
In contrast, for the iLLG and os-LLG equation, the \textit{full} probability densities are $P(\mathbf{m},\mathbf{v},t)$ and $P(\mathbf{m},\mathbf{v},\mathbf{w},t)$, respectively. 
By integrating over the additional variables (i.e. $\mathbf{v}$ for the iLLG, $\mathbf{v}$ and $\mathbf{w}$ for the os-LLG equation), one recovers the marginal probability density $p(\mathbf{m},t)$.
In App.~\ref{app:iLLG_osLLG_KL}, we show that the entropy productions rates of these marginal probability densities can become negative and violate the contractivity.

The second part of this Appendix (App.~\ref{app:coarse_graining} to App.~\ref{app:mag_dynamics}) discusses further details of the numerical results shown in Sec.~\ref{sec:num_results} of the main text.
In App.~\ref{app:coarse_graining} we show the influence of the coarse-graining (binning) on the relative entropy.
App.~\ref{app:gibbs} is a technical statement about the equilibrium Gibbs distribution. 
In App.~\ref{app:Markovlimit}, we look at the Markov limit of the os-LLG equation and compare its dynamics, relative entropy, and entropy production rate to the LLG equation. 
For completeness, we plot the averaged magnetization dynamics of the three magnetization dynamics equations for both parameter configurations of the main text, see App.~\ref{app:mag_dynamics}.
We conclude by a detailed comparison of further non-Markovianity measures in App.~\ref{app:measures}.

\section{Monotonically decreasing relative entropy for LLG equation}
\label{app:cont_LLG}
\subsection{Fokker-Planck of the LLG equation}
\label{app:FPE_LLG}
To show that the LLG Eq.~\eqref{eq:classic-LLG} leads to a contractive dynamics, we use its Fokker-Planck equation (FP) description. 
In contrast to the Langevin equation, the FP equation describes the evolution of the probability distribution of the magnetization at each time $p(\mathbf{m},t)$, rather than the individual trajectory. 
The FP equation of the LLG equation is given as~\cite{Garca_Palacios_1998}
\begin{equation}
    \frac{\partial p(\mathbf{m},t)}{\partial t} = -\nabla_{S^2} \cdot\left(\mathbf{A}(\mathbf{m}) p(\mathbf{m},t) \right) + \frac{1}{2}\nabla_{S^2}\cdot\left(\mathcal{D}(\mathbf{m})\nabla_{S^2}\,p(\mathbf{m},t) \right),
    \label{eq:FPE}
\end{equation}
where the drift vector field is given by $\mathbf{A}(\mathbf{m}) = - \gamma \mathbf{m}\times\mathbf{B}_{\mathrm{eff}} - \gamma\alpha\mathbf{m}\times\left(\mathbf{m}\times\mathbf{B}_{\mathrm{eff}}\right)$, which can be split into a reversible $\mathbf{A}_{\mathrm{rev}}(\mathbf{m}) = -\gamma\mathbf{m}\times\mathbf{B}_{\mathrm{eff}}$ and a irreversible part $\mathbf{A}_{\mathrm{irr}}(\mathbf{m}) = -\gamma\alpha\mathbf{m}\times\left(\mathbf{m}\times\mathbf{B}_{\mathrm{eff}} \right)$.
Further, the diffusion tensor is given by $\mathcal{D}(\mathbf{m}) = 2\alpha k_{\mathrm{B}} T(\mathbb{1} - \mathbf{m}\mathbf{m}^{\mathrm{T}})/\gamma $, whereby the projection operator $\mathcal{P} = \mathbb{1}-\mathbf{m}\mathbf{m}^{\mathrm{T}}$ is equivalent to $\mathcal{P} = \mathbf{m}\times\left(\mathbf{m}\times\mathbf{B}_{\mathrm{eff}}\right)$.

The stationary solution $p(\mathbf{m},t) = p(\mathbf{m})$ is given by the Gibbs distribution $\pi_\beta(\mathbf{m}) = \exp(-\beta\mathcal{H}_{\mathrm{S}}(\mathbf{m}))/\mathcal{Z}$, with the system energy $\mathcal{H}_{\mathrm{S}}(\mathbf{m}) = \mathbf{B}_{\mathrm{eff}}\cdot\mathbf{m}$ and the partition function $\mathcal{Z}$. For further remarks about the stationary distribution see App.~\ref{app:gibbs}.
To show that the Gibbs distribution is the stationary solution to the FP Eq.~\eqref{eq:FPE} we have
\begin{equation}
    \nabla_{S^2}\cdot\left(\mathbf{A}(\mathbf{m})\pi_\beta(\mathbf{m}) \right) = \frac{1}{2}\nabla_{S^2}\cdot\left(\mathcal{D}(\mathbf{m})\nabla_{S^2}\,\pi_\beta(\mathbf{m}) \right).
    \label{eq:detailedbalance}
\end{equation}
This can be rewritten as 
\begin{equation}
    \mathcal{D}(\mathbf{m})\cdot\nabla_{S^2}\pi_\beta(\mathbf{m}) = \mathcal{D}(\mathbf{m})\pi_\beta(\mathbf{m})\cdot\nabla_{S^2}\ln\left(\pi_\beta(\mathbf{m})\right) = \mathcal{D}(\mathbf{m})\pi_\beta(\mathbf{m})\left(-\beta\nabla_{S^2}\mathcal{H}\right) = -\beta \pi_\beta(\mathbf{m})\mathcal{D}(\mathbf{m})\mathbf{B}_{\mathrm{eff}},
\end{equation}
which ultimately reduces to $-\pi_\beta(\mathbf{m})\left(\gamma\alpha\mathbf{m}\times(\mathbf{m}\times\mathbf{B}_{\mathrm{eff}}) \right)$.
The left-hand side of Eq.~\eqref{eq:detailedbalance} is of course given by $\mathbf{A}(\mathbf{m}) = -\pi_\beta(\mathbf{m})\left(\gamma\mathbf{m}\times\mathbf{B}_{\mathrm{eff}} + \gamma\alpha\mathbf{m}\times(\mathbf{m}\times\mathbf{B}_{\mathrm{eff}}) \right) = - \pi_\beta(\mathbf{m})\left( \mathbf{A}_{\mathrm{rev}}(\mathbf{m}) + \mathbf{A}_\mathrm{irr}(\mathbf{m}) \right)$, where the reversible part corresponds to the precession of magnetization in the effective magnetic field, i.e. $\mathbf{A}_{\mathrm{rev}}(\mathbf{m}) = \gamma\mathbf{m}\times\mathbf{B}_{\mathrm{eff}}$, and has $\nabla_{S^2}\cdot\mathbf{A}_{\mathrm{rev}}(\mathbf{m}) = 0$.
Therefore, we obtain 
\begin{equation}
    \left(\mathbf{A}_{\mathrm{irr}}(\mathbf{m})\pi_\beta(\mathbf{m})\right) = \frac{1}{2} \left(\mathcal{D}(\mathbf{m})\nabla_{S^2} \pi_\beta(\mathbf{m})\right),
    \label{eq:detailedbalance_new}
\end{equation}
and hence Eq.~\eqref{eq:FPE} satisfies detailed balance.
This will be employed below to show that the KL measure always monotonically decreases for the LLG equation. 
In fact, we will use a slightly different version of Eq.~\eqref{eq:detailedbalance_new} by noticing that $\nabla_{S^2} \pi_\beta(\mathbf{m}) = \pi_\beta(\mathbf{m})\nabla_{S^2}(-\beta\mathcal{H}_{\mathrm{S}}(\mathbf{m})) = \pi_\beta(\mathbf{m})\nabla_{S^2}\ln\pi_\beta(\mathbf{m})$, such that we get $(\mathbf{A}_{\mathrm{irr}}(\mathbf{m}) -\mathcal{D}(\mathbf{m})\nabla_{S^2}\ln\pi_\beta(\mathbf{m}) )\pi_\beta(\mathbf{m}) = 0$.

\subsection{Monotonically decreasing relative entropy for LLG equation} 
\label{app:markov-LLG}
In the second step, we show that the relative entropy of the LLG Eq.~\eqref{eq:classic-LLG} is monotonically decreasing, i.e. $\dot{D}[p(\mathbf{m},t)||\pi_\beta(\mathbf{m})] \leq 0$ or in terms of the entropy production rate, $\dot{\Sigma}(t)/k_{\mathrm{B}}\geq 0$ (compare with definitions in Sec.~\ref{sec:intro}).
We define the relative entropy as (compare with Eq.~\eqref{eq:D_KL_derivative_short} in the main text)
\begin{equation}
    D\left[ p(\mathbf{m},t) || \pi_\beta(\mathbf{m})\right] = \int_{S^2}p(\mathbf{m},t)\ln\frac{p(\mathbf{m},t)}{\pi_\beta(\mathbf{m})}\mathrm{d}\Omega,
\end{equation}
where $\int_{S^2}\mathrm{d}\Omega$ defines the integral over the sphere surface with $\mathrm{d}\Omega = \sin\vartheta\cdot\mathrm{d}\varphi\mathrm{d}\vartheta$.
The probability distribution $p(\mathbf{m},t)$ describes the probability to find the magnetization $\mathbf{m}$ at time $t$ with $ \int_{S^2}p(\mathbf{m},t)\mathrm{d}\Omega = 1$.
We calculate the change in the relative entropy as
\begin{equation}
    \frac{\mathrm{d}D(t)}{\mathrm{d}t} = \int_{S^2}\frac{\mathrm{d}}{\mathrm{d}t}\left[ p(\mathbf{m},t)\ln\frac{p(\mathbf{m},t)}{\pi_\beta(\mathbf{m})} \right] = \int_{S^2}\left[1 + \ln\frac{p(\mathbf{m},t)}{\pi_\beta(\mathbf{m})} \right]\frac{\partial p(\mathbf{m},t)}{\partial t} \mathrm{d}\Omega = \int_{S^2} \ln\frac{p(\mathbf{m},t)}{\pi_\beta(\mathbf{m})}\frac{\partial p(\mathbf{m},t)}{\partial t}\mathrm{d}\Omega.
    \label{eq:D_KL_derivative}
\end{equation}
Now, we plug the FP Eq.~\eqref{eq:FPE} into the integral expression on the right and obtain
\begin{equation}
    \frac{\mathrm{d}D(t)}{\mathrm{d}t} = -\int_{S^2} \ln\frac{p(\mathbf{m},t)}{{\pi_\beta}(\mathbf{m})}\nabla_{S^2}\cdot\left(\mathbf{A}(\mathbf{m})p(\mathbf{m},t) \right) \mathrm{d}\Omega + \frac{1}{2}\int_{S^2}\ln\frac{p(\mathbf{m},t)}{\pi_\beta(\mathbf{m})} \nabla_{S^2}\cdot\left(\mathcal{D}(\mathbf{m})\nabla_{S^2}\,p(\mathbf{m},t) \right)\mathrm{d}\Omega.
    \label{eq:DLwithFPE}
\end{equation}
For the first term we use the product rule for a scalar function $\phi$ and a vector field $\mathbf{F}$, $\nabla\cdot\left( \phi\mathbf{F} \right) = \left(\nabla \phi \right)\cdot\mathbf{F} + \phi\left( \nabla\cdot \mathbf{F} \right)$. 
Thus, we rewrite the integrand as 
\begin{equation}
    -\ln\frac{p(\mathbf{m},t)}{\pi_\beta(\mathbf{m})} \nabla_{S^2}\cdot\left(\mathbf{A}(\mathbf{m})p(\mathbf{m},t) \right) = -\nabla_{S^2}\cdot\left(\ln\frac{p(\mathbf{m},t)}{\pi_\beta(\mathbf{m})} \left(\mathbf{A}(\mathbf{m}) p(\mathbf{m},t) \right)\right) + \left(\mathbf{A}(\mathbf{m})p(\mathbf{m},t) \right)\cdot\nabla_{S^2}\ln\frac{p(\mathbf{m},t)}{\pi_\beta(\mathbf{m})}.
    \label{eq:first_term_one}
\end{equation}
Using that the integral of the divergence of any tangential vector field over the entire sphere is zero~\cite{Frankel_2011}, the first term in Eq.~\eqref{eq:first_term_one} vanishes, i.e.
\begin{equation}
    -\int_{S^2}\nabla_{S^2}\cdot\left(\ln\frac{p(\mathbf{m},t)}{\pi_\beta(\mathbf{m})} \left(\mathbf{A}(\mathbf{m}) p(\mathbf{m},t) \right) \right)\mathrm{d}\Omega = 0,
\end{equation}
since by construction $\mathbf{A}(\mathbf{m})$ is orthogonal to $\mathbf{m}$.
Therefore, the first term of Eq.~\eqref{eq:DLwithFPE} reduces to 
\begin{equation}
    \int_{S^2} \mathbf{A}(\mathbf{m})p(\mathbf{m},t)\cdot\nabla_{S^2}\ln\frac{p(\mathbf{m},t)}{\pi_\beta(\mathbf{m})}\mathrm{d}\Omega.
\end{equation}
For the second term in Eq.~\eqref{eq:DLwithFPE} we again employ the product rule, such that we can rewrite the integrand as
\begin{equation}
    \ln\frac{p(\mathbf{m},t)}{\pi_\beta(\mathbf{m})}\nabla_{S^2}\cdot \left(\mathcal{D}(\mathbf{m})\nabla_{S^2} \,p(\mathbf{m},t) \right) = \nabla_{S^2} \cdot \left( \ln\frac{p(\mathbf{m},t)}{\pi_\beta(\mathbf{m})} \left(\mathcal{D}(\mathbf{m})\nabla_{S^2}\, p(\mathbf{m},t) \right) \right) - \left( \mathcal{D}(\mathbf{m})\nabla_{S^2}\, p(\mathbf{m},t)\right)\cdot \nabla_{S^2}\, \frac{p(\mathbf{m},t)}{\pi_\beta(\mathbf{m})}.
    \label{eq:intermediate_step}
\end{equation}
Once again we drop the divergence term $\int_{S^2}\nabla_{S^2}\cdot(\ln(p(\mathbf{m},t)/\pi_\beta(\mathbf{m}))(\mathcal{D}(\mathbf{m})p(\mathbf{m},t)))\mathrm{d}\Omega = 0$, since $\mathcal{D}(\mathbf{m})$ is orthogonal to $\mathbf{m}$ and tangential to the sphere~\cite{Frankel_2011}. 
Thus, Eq.~\eqref{eq:intermediate_step} reduces to 
\begin{equation}
    -\int_{S^2}\left( \mathcal{D}(\mathbf{m})\nabla_{S^2}\, p(\mathbf{m},t) \right) \cdot \nabla_{S^2} \ln \frac{p(\mathbf{m},t)}{\pi_\beta(\mathbf{m})}\mathrm{d}\Omega.
    \label{eq:sec_integrand}
\end{equation}
Next, we rewrite this term by using $\nabla\ln p = 1/p\nabla p$ and $\ln p = \ln p/\pi + \ln \pi$. 
This yields 
\begin{equation}
    \nabla_{S^2}\, p(\mathbf{m},t) = p(\mathbf{m},t)\nabla_{S^2} \ln p(\mathbf{m},t) = p(\mathbf{m},t) \left[\nabla_{S^2}\ln\frac{p(\mathbf{m},t)}{\pi_\beta(\mathbf{m})} + \nabla_{S^2} \ln\pi_\beta(\mathbf{m}) \right],
\end{equation}
and, by changing the order of multiplication, Eq.~\eqref{eq:sec_integrand} becomes
\begin{align}
    -\int_{S^2}p(\mathbf{m},t)\left(\nabla_{S^2}\,\ln\frac{p(\mathbf{m},t)}{\pi_\beta(\mathbf{m})} \right)^{\mathrm{T}}&\mathcal{D}(\mathbf{m})\left(\nabla_{S^2} \ln \frac{p(\mathbf{m},t)}{\pi_\beta(\mathbf{m})}\right)\mathrm{d}\Omega \nonumber\\
    &- \int_{S^2}p(\mathbf{m},t)\left(\nabla_{S^2} \ln \pi_\beta(\mathbf{m}) \right)^{\mathrm{T}}\mathcal{D}(\mathbf{m})\left( \nabla_{S^2} \ln \frac{p(\mathbf{m},t)}{\pi_\beta(\mathbf{m})} \right)\mathrm{d}\Omega.
\end{align}
Finally, we recombine these terms and obtain
\begin{equation}
    \int_{S^2}p(\mathbf{m},t)\nabla_{S^2}\ln\frac{p(\mathbf{m},t)}{\pi_\beta(\mathbf{m})}\cdot\left(\mathbf{A}_{\mathrm{rev}}(\mathbf{m}) + \mathbf{A}_{\mathrm{irr}}(\mathbf{m})-\frac{1}{2}\mathcal{D}(\mathbf{m})\nabla_{S^2}\ln\pi_\beta(\mathbf{m}) \right)\mathrm{d}\Omega.
    \label{eq:termfull}
\end{equation}
We find that the first term in the sum in the integral (i.e. integral over $\mathbf{A}_{\mathrm{rev}}(\mathbf{m})$) vanishes. To show this we use $\nabla\ln(p(\mathbf{m},t)/\pi_\beta(\mathbf{m})) = \nabla p(\mathbf{m},t)/p(\mathbf{m},t) - \nabla \pi_\beta(\mathbf{m})/\pi_\beta(\mathbf{m})$, and that $\nabla_{S^2}\cdot\mathbf{A}_{\mathrm{rev}}(\mathbf{m}) = 0$ and $\nabla_{S^2}\cdot(\mathbf{A}_{\mathrm{rev}}(\mathbf{m})\pi_\beta(\mathbf{m})) = 0$. 
Hence, we get
\begin{align}
    \int_{S^2} p(\mathbf{m},t)&\left(\frac{\nabla_{S^2}\, p(\mathbf{m},t)}{p(\mathbf{m},t)} - \frac{\nabla_{S^2}\, \pi_\beta(\mathbf{m})}{\pi_\beta(\mathbf{m})} \right)\cdot \mathbf{A}_{\mathrm{rev}}(\mathbf{m})\mathrm{d}\Omega \nonumber\\
    &= \underbrace{\int_{S^2}\nabla_{S^2}\,p(\mathbf{m},t)\cdot \mathbf{A}_{\mathrm{rev}}(\mathbf{m})\mathrm{d}\Omega}_{\substack{(1)}} - \underbrace{\int_{S^2}\frac{p(\mathbf{m},t)}{\pi_\beta(\mathbf{m})}\nabla_{S^2}\, \pi_\beta(\mathbf{m})\cdot \mathbf{A}_{\mathrm{rev}}(\mathbf{m})\mathrm{d}\Omega}_{\substack{(2)}},
    \label{eq:second_term}
\end{align}
where (1) can be rewritten by the product rule to be 
\begin{equation}
    \int_{S^2}\nabla_{S^2}\, p(\mathbf{m},t)\cdot \mathbf{A}_{\mathrm{rev}}(\mathbf{m})\mathrm{d}\Omega = \int_{S^2}\nabla_{S^2}\cdot\left(p(\mathbf{m},t)\mathbf{A}_{\mathrm{rev}}(\mathbf{m}) \right)\mathrm{d}\Omega - \int_{S^2}p(\mathbf{m},t)\underbrace{\nabla_{S^2}\cdot\mathbf{A}_{\mathrm{rev}}(\mathbf{m})}_{\substack{= 0}}\mathrm{d}\Omega = 0,
\end{equation}
with the first term on the right vanishing as the integral of the divergence term of $\mathbf{A}_{\mathrm{rev}}(\mathbf{m})$ on $S^2$ is again zero. 

The second term in Eq.~\eqref{eq:second_term} can be shown to vanish by observing that $\nabla_{S^2}\cdot(\mathbf{A}_{\mathrm{rev}}(\mathbf{m})\pi_\beta(\mathbf{m})) = \pi_\beta(\mathbf{m})\nabla_{S^2} \cdot \mathbf{A}_{\mathrm{rev}}(\mathbf{m}) + \mathbf{A}_{\mathrm{rev}}(\mathbf{m})\cdot\nabla_{S^2}\,\pi_\beta(\mathbf{m})$, where the term on the left and the first term on the right vanish. 
Hence, the second term on the right also needs to vanish, i.e. $\mathbf{A}_{\mathrm{rev}}(\mathbf{m})\cdot\pi_\beta(\mathbf{m}) = 0$. 
Therefore, Eq.~\eqref{eq:termfull} reduces to 
\begin{equation}
    \int_{S^2}p(\mathbf{m},t)\nabla_{S^2}\ln\frac{p(\mathbf{m},t)}{\pi_\beta(\mathbf{m})}\cdot\left( \mathbf{A}_{\mathrm{irr}}(\mathbf{m})-\frac{1}{2}\mathcal{D}(\mathbf{m})\nabla_{S^2}\ln\pi_\beta(\mathbf{m}) \right)\mathrm{d}\Omega.
\end{equation}
However, this vanishes due to the detailed balance condition discussed in App.~\ref{app:FPE_LLG}.
So, we are left with the following expression for the change in the relative entropy,
\begin{equation}
    \frac{\mathrm{d}D(t)}{\mathrm{d}t} = - \frac{1}{2}\int_{S^2} p(\mathbf{m},t)\left( \nabla_{S^2} \ln\frac{p(\mathbf{m},t)}{\pi_\beta(\mathbf{m})} \right)^{\mathrm{T}}\mathcal{D}(\mathbf{m})\left(\nabla_{S^2}\ln\frac{p(\mathbf{m},t)}{\pi_\beta(\mathbf{m})}\right)\mathrm{d}\Omega.
    \label{eq:DKL_final_LLG}
\end{equation}
We define $\nabla_{S^2}\ln (p(\mathbf{m},t)/\pi_\beta(\mathbf{m})) = \mathbf{p}$ and find that $\mathbf{p}\perp \mathbf{m}$, where we use the fact that the surface gradient $\nabla_{S^2}$ is always tangent to the sphere and the sphere’s normal at point $\mathbf{m}$ is $\mathbf{m}$, so the surface gradient is perpendicular to $\mathbf{m}$.
Therefore, $\mathbf{p}^{\mathrm{T}}\mathcal{D}(\mathbf{m})\mathbf{p} = \mathbf{p}^{\mathrm{T}}(\mathbb{1}-\mathbf{m}\mathbf{m}^{\mathrm{T}})\mathbf{p} = \abs{\mathbf{p}}^2 \geq 0$. 
Further, by definition $p(\mathbf{m},t) \geq 0\,\forall t$, such that we finally have $\dot{D}(t)\leq 0$ or rather $\dot{\Sigma}(t)/k_{\mathrm{B}} \geq 0$.  
This concludes the proof of the contractivity of the LLG equation.

\section{Non-monotonic decrease of the relative entropy for the iLLG  and os-LLG equation}
\label{app:iLLG_osLLG_KL}
After we have shown above that the relative entropy monotonically decreases for the LLG Eq.~\eqref{eq:classic-LLG}, we would like to show that this is no longer possible to show for the iLLG Eq.~\eqref{eq:inertial-LLG} as well as for the os-LLG Eq.~\eqref{eq:open-system LLG}.

\subsection{iLLG equation}
We start by noting that the inertial LLG Eq.~\eqref{eq:inertial-LLG} can be rewritten as a set of two coupled first-order differential equations~\cite{Mondal_2021}
\begin{align}
    \frac{\mathrm{d}\mathbf{m}}{\mathrm{d}t} &= \mathbf{v}, \label{eq:embedding_iLLG1} \\
    \frac{\mathrm{d}\mathbf{v}}{\mathrm{d}t} &= -\frac{\gamma}{\eta}\mathbf{m}\times\left(\mathbf{m}\times\mathbf{B}_{\mathrm{eff}}\right) - \frac{\gamma}{\eta}\mathbf{m}\times\left(\mathbf{m}\times\mathbf{b}(t) \right) - \frac{\alpha}{\eta}\mathbf{v}-\frac{1}{\eta}\mathbf{m}\times\mathbf{v} -\mathbf{m}\abs{\mathbf{v}}^2.
    \label{eq:embedding_iLLG2}
\end{align}
To follow the same logic as in App.~\ref{app:cont_LLG}, we set up the corresponding Kramer-type Fokker-Planck equation for a probability distribution $P(\mathbf{m},\mathbf{v},t) = P(\mathbf{X},t)$ with $\mathbf{X} = (\mathbf{m},\mathbf{v})$.
First we notice that Eq.~\eqref{eq:embedding_iLLG2} can be written as a set of stochastic differential equations in Stratonovich form, as $\mathrm{d}X^{\alpha} = A^{\alpha}(X)\mathrm{d}t + B_i^{\alpha}(X)\circ\mathrm{d}W_t^i$, where $\alpha = m,v$ refers to the differential equation for $\mathbf{m}$ and $\mathbf{v}$, respectively~\cite{arnold1974stochastic}. The index $i$ denotes the different independent noise sources (for the iLLG case $i=3$, as $\mathbf{b}(t)$ is three-dimensional Cartesian vector with independent components).
Then the corresponding Fokker-Planck equation is defined as 
\begin{equation}
    \partial_t P(\mathbf{X},t) = \mathcal{L}_{\mathrm{FP}}P(\mathbf{X},t) = \partial_\alpha (A^\alpha P(\mathbf{X},t)) + \frac{1}{2}\partial_\alpha\partial_\beta(\mathcal{D}^{\alpha\beta}P(\mathbf{X},t)),
\end{equation}
with the Fokker-Planck operator $\mathcal{L}_{\mathrm{FP}}$ and the diffusion tensor being defined as $\mathcal{D}^{\alpha\beta} = \sum_i B_i^{\alpha}\otimes B_i^\beta$. 
From Eqs.~\eqref{eq:embedding_iLLG1} and~\eqref{eq:embedding_iLLG2} we can read off the following drift coefficients, $\mathbf{A}^m = \mathbf{v}$ and $\mathbf{A}^v = F_{\mathrm{det}}(\mathbf{m},\mathbf{v})$, where 
\begin{equation}
    F_{\mathrm{det}}(\mathbf{m},\mathbf{v}) = -\frac{\gamma}{\eta}\mathbf{m}\times\left(\mathbf{m}\times\mathbf{B}_{\mathrm{eff}}\right) - \frac{\alpha}{\eta} \mathbf{v} -\frac{1}{\eta}\mathbf{m}\times\mathbf{v} - \mathbf{m}\abs{\mathbf{v}}^2
\end{equation}
is the deterministic part of the right-hand side of Eq.~\eqref{eq:embedding_iLLG2}.
Since the noise only acts in Eq.~\eqref{eq:embedding_iLLG2} we find that $B_i^{m} = 0$ and $B_i^v = -\gamma\xi\eta^{-1}\mathbf{m}\times(\mathbf{m}\times\mathbf{e}_i)$ with $\xi = \sqrt{2\alpha k_{\mathrm{B}}T/\gamma}$, arising from the strength of the noise fluctuations.
This leads to $\mathcal{D}^{vv}(\mathbf{m}) = 2\alpha k_{\mathrm{B}}T\gamma/\eta^2\,(\mathbb{1} - \mathbf{m}\mathbf{m}^{\mathrm{T}})$, which leads to an isotropic diffusion of the ``velocity'' variable in the two-dimensional tangent plane. 
In addition, all diffusion terms with the index $m$ disappear, i.e. $\mathcal{D}^{mm} = \mathcal{D}^{mv} = \mathcal{D}^{vm} = 0$.
Finally, we arrive at the Fokker-Planck equation for the probability distribution $P(\mathbf{m},\mathbf{v},t)$,
\begin{equation}
    \frac{\partial P(\mathbf{m},\mathbf{v},t)}{\partial t} = -\nabla_{\mathbf{m}}\cdot\left( \mathbf{v} P(\mathbf{m},\mathbf{v},t) \right) - \nabla_{\mathbf{v}}\cdot\left( \mathbf{F}_{\mathrm{det}}(\mathbf{m},\mathbf{v}) P(\mathbf{m},\mathbf{v},t)\right) + \frac{1}{2}\nabla^2_{\mathbf{v}}\left(\mathcal{D}^{vv}(\mathbf{m}) P(\mathbf{m},\mathbf{v},t)\right)
    \label{eq:FPE_iLLG}
\end{equation}
We notice that $\nabla_{\mathbf{v}}^2\mathcal
D^{vv}(\mathbf{m}) = 0$ and that $\mathcal{D}^{vv}(\mathbf{m})$ can be expressed in the orthonormal basis of the tangent plane $\{v_1,v_2\}$, such that $\mathcal{D}^{vv}_{ij}(\mathbf{m})/2 = D_{v}\delta_{ij} = \alpha k_{\mathrm{B}}T\gamma/\eta^2$. Here, with absorbed the prefactor $1/2$ into the definition of $D_v$. 
Therefore, the Fokker-Planck equation becomes
\begin{equation}
    \frac{\partial P(\mathbf{m},\mathbf{v},t)}{\partial t} = -\nabla_{\mathbf{m}}\cdot\left( \mathbf{v} P(\mathbf{m},\mathbf{v},t) \right) - \nabla_{\mathbf{v}}\cdot\left( \mathbf{F}_{\mathrm{det}}(\mathbf{m},\mathbf{v}) P(\mathbf{m},\mathbf{v},t)\right) + \mathcal{D}_{v}\nabla^2_{\mathbf{v}}(\mathbf{m}) P(\mathbf{m},\mathbf{v},t).
    \label{eq:FPE_iLLG2}
\end{equation}
For the total probability distribution of $P(\mathbf{m},\mathbf{v},t)$ it still holds that $\dot{D}(t) \leq 0$, since the full probability distribution of $\mathbf{m}$ and $\mathbf{v}$ is known. There are no hidden degrees of freedom and the dynamics is Markovian.
This, however, no longer holds if we move to a description of the marginal probability density $p(\mathbf{m},t)$, which we define as 
\begin{equation}
    p(\mathbf{m},t) = \int P(\mathbf{m},\mathbf{v},t)\mathrm{d}\mathbf{v},
\end{equation}
where we integrate over the full ``velocity'' subspace $\{v_1,v_2\}\in\mathbb{R}^2$ and $\mathrm{d}\mathbf{v} = \mathrm{d}v_1\mathrm{d}v_2$.
In order to achieve this, we employ the Mori-Zwanzig formalism (see e.g. Ref.~\cite{Vrugt_2020,Breuer_2007}) to obtain the marginal probability distribution $p(\mathbf{m},t)$.
Therefore, we define a projection operator $\mathcal{P}$, such that 
\begin{equation}
    \left(\mathcal{P}P\right)(\mathbf{m},\mathbf{v},t) = \varphi(\mathbf{v}|\mathbf{m})\int P(\mathbf{m},\mathbf{v}',t)\,\mathrm{d}\mathbf{v}',
\end{equation}
where for each fixed $\mathbf{m}$, $\varphi(\mathbf{v}|\mathbf{m})$ is a probability density in $\mathbf{v}$, with $\int\varphi(\mathbf{v}|\mathbf{m})\,\mathrm{d}\mathbf{v} =1$.
We can easily check that $\mathcal{P}$ leaves the marginal probability density $p(\mathbf{m},t)$ unchanged
\begin{equation}
    \int \left(\mathcal{P}P\right)(\mathbf{m},\mathbf{v},t)\,\mathrm{d\mathbf{v}} = \int \varphi(\mathbf{v}|\mathbf{m})\int P(\mathbf{m},\mathbf{v}',t)\,\mathrm{d}\mathbf{v}'\,\mathrm{d}\mathbf{v} = \int P(\mathbf{m},\mathbf{v}',t)\,\mathrm{d}\mathbf{v}',
\end{equation}
and that $\mathcal{P}^2 = \mathcal{P}$,
\begin{equation}
    \mathcal{P}\left(\mathcal{P}P(\mathbf{m},\mathbf{v},t)\right)(\mathbf{m},\mathbf{v},t) = \varphi(\mathbf{v}|\mathbf{m})\int \varphi(\mathbf{v}'|\mathbf{m}) \left( \int P(\mathbf{m},\mathbf{v}'',t)\,\mathrm{d}\mathbf{v}'' \right)\,\mathrm{d}\mathbf{v}' = \varphi(\mathbf{v}|\mathbf{m})\int P(\mathbf{m},\mathbf{v}'',t)\,\mathrm{d}\mathbf{v}''.
\end{equation}
The complement operator is defined as $\mathcal{Q} = 1-\mathcal{P}$, such that $\mathcal{P}\mathcal{Q} = \mathcal{Q}\mathcal{P} =0$.
Next, we apply the projection operators $\mathcal{P}$ and 
$\mathcal{Q}$ to the FPE equation
\begin{align}
    \partial_t\left(\mathcal{P}P\right)(\mathbf{m},\mathbf{v},t) &= \mathcal{P}\mathcal{L}_{\mathrm{FP}}\left(\mathcal{P}P(\mathbf{m},\mathbf{v},t) + \mathcal{Q}P(\mathbf{m},\mathbf{v},t)\right), \\
    \partial_t \left(\mathcal{Q}P\right)(\mathbf{m},\mathbf{v},t) &= \mathcal{Q}\mathcal{L}_{\mathrm{FP}}\left( \mathcal{P}P(\mathbf{m},\mathbf{v},t) + \mathcal{Q}P(\mathbf{m},\mathbf{v},t)\right).
\end{align}
We can make use of linearity and $\mathcal{P}\mathcal{Q} = 0$ and arrive at
\begin{align}
    \partial_t \left(\mathcal{P}P \right)(\mathbf{m},\mathbf{v},t) &= \mathcal{P}\mathcal{L}_{\mathrm{FP}}\mathcal{P}P(\mathbf{m},\mathbf{v},t) + \mathcal{P}\mathcal{L}_{\mathrm{FP}}\mathcal{Q}P(\mathbf{m},\mathbf{v},t),\label{eq:P_eq}\\
    \partial_t \left( \mathcal{Q} P\right)(\mathbf{m},\mathbf{v},t) &= \mathcal{Q} \mathcal{L}_{\mathrm{FP}}\mathcal{P}P(\mathbf{m},\mathbf{v},t) + \mathcal{Q}\mathcal{L}_{\mathrm{FP}}\mathcal{Q}P(\mathbf{m},\mathbf{v},t).
    \label{eq:Q_eq}
\end{align}
The formal solution of Eq.~\eqref{eq:Q_eq} is given by~\cite{Vrugt_2020}
\begin{equation}
    \mathcal{Q}P(\mathbf{m},\mathbf{v},t) = e^{\mathcal{Q}\mathcal{L}_{\mathrm{FP}} t}\mathcal{Q}P(\mathbf{m},\mathbf{v},t)+\int_0^t\mathrm{d}s\, e^{\mathcal{Q}\mathcal{L}_{\mathrm{FP}}(t-s)}\mathcal{Q}\mathcal{L}_{\mathrm{FP}}\mathcal{P}P(\mathbf{m},\mathbf{v},t),
\end{equation}
which can be substituted into Eq.~\eqref{eq:P_eq} to obtain the Nakajima-Zwanzig equation
\begin{equation}
    \partial_t \left( \mathcal{P}P \right)(\mathbf{m},\mathbf{v},t) = \mathcal{P}\mathcal{L}_{\mathrm{FP}}\mathcal{P}P(\mathbf{m},\mathbf{v},t) + \mathcal{P}\mathcal{L}_{\mathrm{FP}}e^{\mathcal{Q}\mathcal{L}_{\mathrm{FP}}t}\mathcal{Q}P(\mathbf{m},\mathbf{v},0) + \int_0^t \mathrm{d}s\,\mathcal{P}\mathcal{L}_{\mathrm{FP}} e^{\mathcal{Q}\mathcal{L}_{\mathrm{FP}}(t-s)}\mathcal{Q}\mathcal{L}_{\mathrm{FP}}\mathcal{P}\left( \mathcal{P}P\right)(\mathbf{m},\mathbf{v},s).
    \label{eq:Nakajima_Zwanzig}
\end{equation}
The first term describes the instantaneous (Markovian) responds of the system. The second term is an initial-condition term depending on $\mathcal{Q}P(\mathbf{m},\mathbf{v},0)$.
The relevant term for the purpose of this derivation is the last term, which convolves the past $(\mathcal{P}P)(\mathbf{m},\mathbf{v},s)$ into the present at time $t$.
The memory kernel is made explicit by defining $\mathcal{K}(t-t') = \mathcal{P}\mathcal{L}_{\mathrm{FP}}e^{\mathcal{Q}\mathcal{L}_{\mathrm{FP}}(t-s)}\mathcal{Q\mathcal{L}_{\mathrm{FP}}}\mathcal{P}$.

Now, we can integrate the Nakajima-Zwanzig Eq.~\eqref{eq:Nakajima_Zwanzig} over $\mathbf{v}$ such that we obtain an equation for the marginal probability density $p(\mathbf{m},t)$.
Therefore, we use the above identity $\int (\mathcal{P}P)(\mathbf{m},\mathbf{v},t)\,\mathrm{d}\mathbf{v} = \int P(\mathbf{m},\mathbf{v},t)\,\mathrm{d}\mathbf{v} = p(\mathbf{m},t)$ and get
\begin{equation}
    \partial_t p(\mathbf{m},t) = \int \mathrm{d}\mathbf{v}\,\partial_t \left( \mathcal{P}P \right)(\mathbf{m},\mathbf{v},t) = \int \mathrm{d}\mathbf{v}\,\left[ \mathcal{P}\mathcal{L}_{\mathrm{FP}}\mathcal{P}P(\mathbf{m},\mathbf{v},t) + \int_0^t\mathrm{d}s\,\mathcal{K}(t-s) \left(\mathcal{P}P \right)(\mathbf{m},\mathbf{v},t) + \mathrm{init.}\right].
\end{equation}
By defining $\mathcal{M}(\mathbf{m},t-s) := \int\mathrm{d}\mathbf{v}\,\mathcal{K}(t-s)\varphi(\mathbf{v}|\mathbf{m})$ the equation simplifies to 
\begin{equation}
    \partial_t p(\mathbf{m},t) = I_{\mathrm{inst.}}\left[p(\mathbf{m},t)\right] + \int_0^t\mathrm{d}s\, \mathcal{M}(\mathbf{m},t-s)p(\mathbf{m},s) + I_{\mathrm{init.}}.
    \label{eq:marginal_density}
\end{equation}
This gives us an expression for the marginal probability density over $\mathbf{m}$ only.
Hence, we can use this expression and plug it into Eq.~\eqref{eq:D_KL_derivative}, such that we obtain
\begin{equation}
    \dot{D}(t) = \underbrace{\int_{S^2}\ln \frac{p(\mathbf{m},t)}{\pi_\beta(\mathbf{m})} I_{\mathrm{inst.}}\left[p(\mathbf{m},t) \right]\,\mathrm{d}\Omega}_{\substack{(\mathrm{A})}} + \underbrace{\int_{S^2} \ln\frac{p(\mathbf{m},t)}{\pi_\beta(\mathbf{m})}\left(\int_0^t\mathrm{d}s\,\mathcal{M}(\mathbf{m},t-s)p(\mathbf{m},s) \right)\mathrm{d}\Omega}_{\substack{\mathrm{(B)}}} + \underbrace{\int_{S^2}\ln \frac{p(\mathbf{m},t)}{\pi_\beta(\mathbf{m})}I_{\mathrm{init.}}\,\mathrm{d}\Omega}_{\substack{\mathrm{(C)}}}.
\end{equation}
The term (A) is like the term from the Fokker-Planck equation of the LLG equation and can be rearranged to positive quadratic form with a negative sign (compare to Eq.~\eqref{eq:DKL_final_LLG}),
\begin{equation}
    \mathrm{(A)} = -\frac{1}{2}\int_{S^2}p(\mathbf{m},t)\left(\nabla_{S^2}\ln\frac{p(\mathbf{m},t)}{\pi_\beta(\mathbf{m})} \right)^{\mathrm{T}}\mathcal{D}_{\mathrm{eff}}(\mathbf{m})\left(\nabla_{S^2} \ln \frac{p(\mathbf{m},t)}{\pi_\beta(\mathbf{m})} \right)\,\mathrm{d}\Omega \leq 0,
\end{equation}
which gives the usual monotone contribution. 
In contrast, the second term (B) mixes the marginal probability density $p(\mathbf{m},t)$ (at the current time $t$) with the marginal probability density $p(\mathbf{m},s)$ at a past time $s$.
This term can not be collapsed to a positive quadratic form and the sign of (B) depends, in general, on the history of the system (is sign-indefinite). 
This contribution can at times ``overwhelm'' the instantaneous part making the conclusion $\dot{D}(t) \leq 0$ incorrect. 
This concludes the proof that the relative entropy $D(t)$ no longer monotonically decreases for the iLLG equation. 
The third term (C) arises from the initial conditions, but its details are not relevant for the conclusion made above. 

\subsection{os-LLG equation}
Similarly, the os-LLG Eq.~\eqref{eq:open-system LLG} can be written as a set of three coupled first-order (Markovian) differential equations~\cite{Anders2022quantum}
\begin{align}
    \frac{\mathrm{d}\mathbf{m}}{\mathrm{d}t} &= \mathbf{m}\times\left(\mathbf{B}_{\mathrm{eff}} + \tilde{\mathbf{b}}(t) + \mathbf{v}  \right) \\
    \frac{\mathrm{d}\mathbf{v}}{\mathrm{d}t} &= \mathbf{w} \\
    \frac{\mathrm{d}\mathbf{w}}{\mathrm{d}t} &= -\omega_0^2\mathbf{v} - \Gamma \mathbf{w} + \alp\mathbf{m}. 
\end{align}
Without explicitly showing this, a similar argument to the one above for the iLLG equation can be constructed for the os-LLG equation.
First, by defining the corresponding FP equation for the \textit{full} probability distribution $P(\mathbf{m},\mathbf{v},\mathbf{w},t)$ and the marginal probability distribution as $p(\mathbf{m},t) = \int\int P(\mathbf{m},\mathbf{v},\mathbf{w},t)\mathrm{d}\mathbf{v}\mathrm{d}\mathbf{w}$. 
Second, the FP description is reduced to the marginal probability distribution $\partial p(\mathbf{m},t)/\partial t$, which gives rise to ``memory'' terms that lead to negative entropy production rates.

\section{Influence of the coarse-graining on the relative entropy}
\label{app:coarse_graining}
\begin{figure*}[t]
    \centering
    \includegraphics[width=0.93\textwidth]{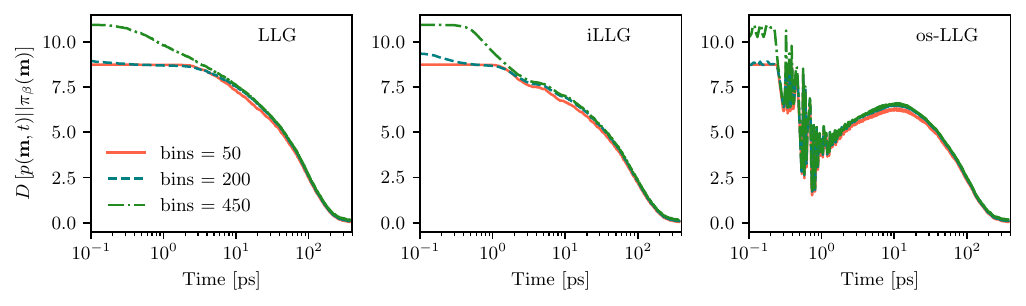}
    \caption{\textbf{Relative entropy for different levels of coarse graining.} Left panel: The relative entropy over time until the system has arrived at equilibrium (i.e. $D(t) = 0$ and $\dot{D}(t) = 0$). The system evolves via the standard LLG equation, but the resolution of the coarse-graining is increased from 50 (orange solid line) to 200 (blue dashed line) to 450 (green dash-dotted line) bins.
    Center panel and right panel: The same as in the left panel, but the system evolves via iLLG and os-LLG equation, respectively.
    The initial condition and the magnetic field are fixed by the parallel parameter configuration. 
    The relative entropy is calculated for $N_{\mathrm{trac}} = 2000$ trajectories.
    }
    \label{fig:DKL_binning}
\end{figure*}

Here, the influence of the coarse-graining on the numerically calculated relative entropy is discussed. 
The relative entropy $D[p(\mathbf{m},t)||\pi_\beta(\mathbf{m})]$ of the LLG, iLLG, and os-LLG equations for a coarse-graining of 50 (orange solid line), 200 (blue dashed line), and 450 (green dash-dotted line) bins (see Fig.~\ref{fig:DKL_binning}) is compared.
The finer the graining, the larger is the initial relative entropy. 
Further, finer coarse-graining is helpful to resolve the inertial motion of the magnetization, however, at the same time increases the statistical noise (middle and right panel in Fig.~\ref{fig:DKL_binning}).  

\begin{figure*}[t]
    \centering
    \includegraphics[width=0.93\textwidth]{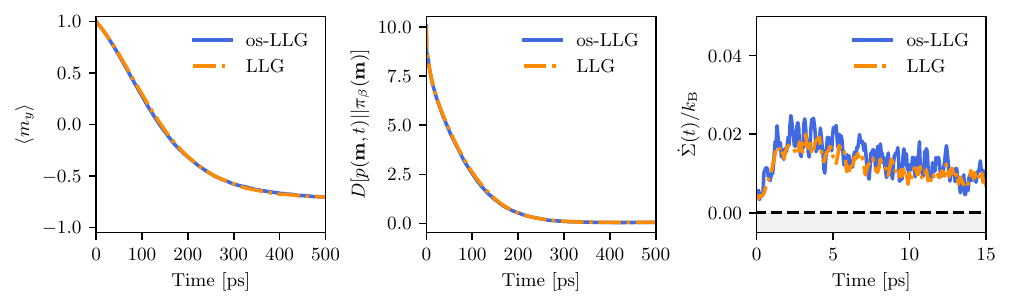}
    \caption{\textbf{Comparison of the LLG equation and the os-LLG equation with ``Ohmic'' parameters.} Here we compare the dynamics, the relative entropy, and the entropy production rate of the LLG equation (dash-dotted orange line) with the Markovian limit of the os-LLG equation (blue solid line). 
    We find agreement between the LLG and the os-LLG equation (in the Markovian limit) for all three quantities. 
    The damping parameter is fixed to $\alpha =0.15$ with the initial state and field strength being fixed by configuration 1 (see main text). The Lorentzian parameters are $\alpha = 125~\mathrm{THz}^3$, $\Gamma = 4.7$ THz, and $\nu_0 = 7.9$ THz.}
    \label{fig:app_Ohmiclimit}
\end{figure*}

\begin{figure*}[b]
    \centering
    \includegraphics[width=0.93\textwidth]{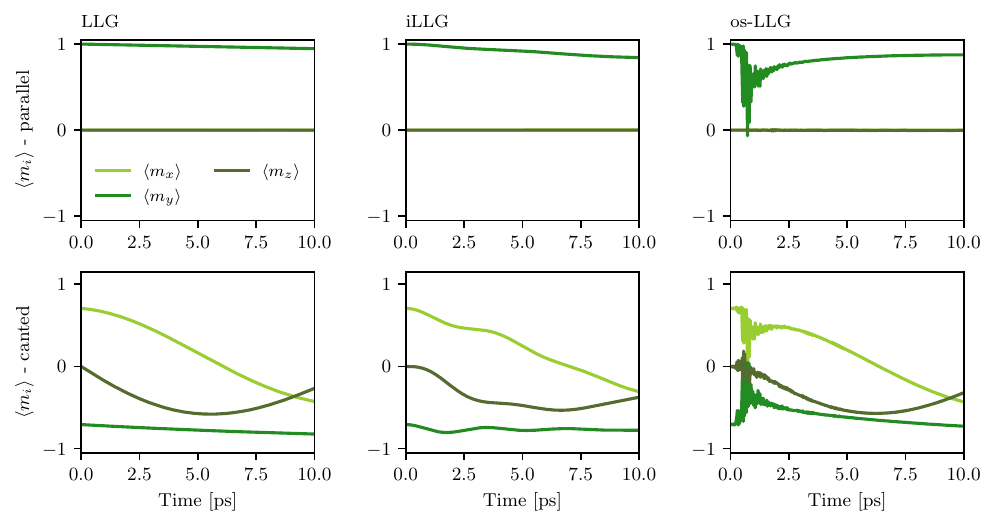}
    \caption{\textbf{Averaged magnetization dynamics for the three equations of motion in Cartesian coordinates.} Left panel: The averaged Cartesian components of the magnetization vector, i.e., light green line for $\langle m_x \rangle$, green line for $\langle m_y \rangle$, and dark green line for $\langle m_z \rangle$ for the dynamics of the LLG Eq.~\eqref{eq:classic-LLG}. Middle panel: The same for the magnetization vector components of the iLLG Eq.~\eqref{eq:inertial-LLG}, which shows an additional faster oscillation on top of the precessional motion. Right panel: The components of the magnetization dynamics of the os-LLG Eq.~\eqref{eq:open-system LLG}, which gives rise to a complex frequency spectrum due to the memory kernel.}
    \label{fig:mag_dyn}
\end{figure*}

\section{Mean Force Gibbs State}
\label{app:gibbs}
Knowing the steady state of the system is important because it is the fixed point of its dynamics and therefore necessary for the calculation of the relative entropy~\cite{strasberg2019non}.
Generally speaking, in strong-coupling thermodynamics~\cite{Seifert_2016, Talkner_2020}, the steady state is not always the Gibbs state $\pi_\beta = e^{-\beta E_\alpha}/\mathcal{Z}$, with 
$\beta$ and $\mathcal{Z} = \sum_\alpha e^{-\beta E_\alpha}$ being the inverse temperature and the partition function, respectively, and mean force corrections to the Gibbs state are relevant~\cite{Trushechkin_2022}.
However, for a classical three-dimensional \textit{isotropic} spin-boson system, i.e., the underlying open system model for the os-LLG Eq.~\eqref{eq:open-system LLG}, the steady state is always given by the Gibbs state $\pi_\beta$.
This holds for any coupling strength~\cite{Hartmann_2023}.
Therefore, we do not need to further specify the coupling strength here (see Fig.~\ref{fig:distributions}).

\section{Markovian limit of the os-LLG equation}
\label{app:Markovlimit}
It has been shown in Ref.~\cite{Anders2022quantum} that Eq.~\eqref{eq:open-system LLG} reduces to the standard LLG equation if the Lorentzian parameters are chosen such that $\nu_0$ and $\Gamma$ are much larger than the spin's typical frequency, i.e. $\nu_0,\,\Gamma \gg \omega_{\mathrm{L}}$.
Thus, the memory kernel term reduces to, $\alpha\,\partial_t\mathbf{M}(t)$, leading to a time-local (Markovian) expression.
We show this numerically in Fig.~\ref{fig:app_Ohmiclimit} for the parallel configuration, where we contrast the LLG equation simulation from the main text with simulations of the os-LLG equation in its Markovian limit. 
The Lorentzian parameters now are $\alpha = 125~\mathrm{THz}^3$, $\Gamma = 4.7$ THz, and $\nu_0 = 7.9$ THz, which again gives a damping of $\eta = \alpha\Gamma/\omega_0^4 = 0.15$.
We find that in this Markovian limit the dynamics of the magnetization matches nicely, see left panel of Fig.~\ref{fig:app_Ohmiclimit}. 
The same is observed for the relative entropy and the entropy production rate, see middle and right panel in Fig,~\ref{fig:app_Ohmiclimit}, respectively. 

\section{Averaged magnetization dynamics}
\label{app:mag_dynamics}
Here, we show the averaged magnetization dynamics of the three equations of motion (see Fig.~\ref{fig:mag_dyn}), whereas the main text discusses quantities that are secondary to the magnetization dynamics of the individual equations, e.g. relative entropy, or its probability distribution (Fig.~\ref{fig:distributions}). 
The parameters are $\alpha =0.15$, $\tau = 120$~fs, and the Lorentzian is parameterized as $\alp = 242\,\mathrm{THz}^3$, $\nu_0 = 4.2$~THz, and $\Gamma = 0.2$~THz.
The initial conditions and the field strength are given by the parallel configuration in the top row and by the canted configuration in the bottom row of Fig.~\ref{fig:mag_dyn}.

Though the motion of the first few pico-second deviates between the three equations of motion, the overall dynamics of the three magnetization vector components remains similar. This is observed for both configurations. 
Due to the initial alignment of the magnetic field along the negative $y$-direction, the $\langle m_x \rangle$ and $\langle m_z \rangle$ start to precess around the magnetic field direction for the canted configuration (see light and dark green lines in bottom panel of Fig.~\ref{fig:mag_dyn}). 

\section{Comparing different measures of non-Markovianity}
\label{app:measures}
\begin{figure*}[b]
    \centering
    \includegraphics[width=0.93\textwidth]{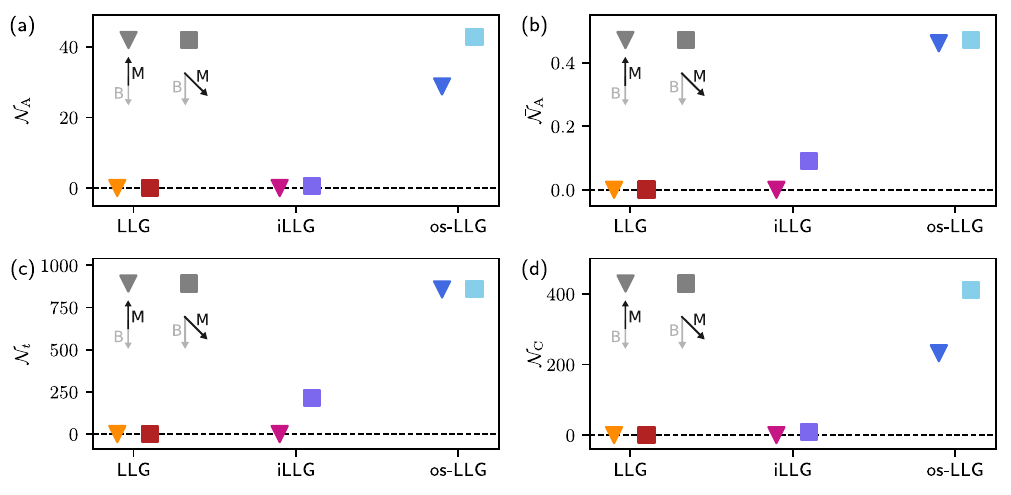}
    \caption{\textbf{Comparison of four different non-Markovianity measures.} (a), (b) The same measures as in Fig.~\ref{fig:EP_measure_area} in the main text. 
    (c) The measure $\mathcal{N}_t$ counts all time steps for which the entropy production rate is negative. (d) The measure counts how often the entropy production rate changes its sign.
    All four measures show the same qualitative agreement. 
    }
    \label{fig:app_measures}
\end{figure*}

In the main text we compared two measures to quantify the amount of non-Markovianity of a process. 
The first measure $\mathcal{N}_{\mathrm{A}}$ (``negative EPR window'') integrates the entropy production rate, whenever it turns negative~\cite{Laine_2010}, compare with Eq.~\eqref{eq:nonMarkov_measure_area}. 
The second measure $\overline{\mathcal{N}}_{\mathrm{A}}$ (``relative EPR window''), see Eq.~\eqref{eq:nonMarkov_measure_area_norm}, is like $\mathcal{N}_{\mathrm{A}}$, but normalized by the total absolute value of the entropy production rate.
They are compared in Fig.~\ref{fig:EP_measure_area} and are discussed in detail in Sec.~\ref{sec:num_results}.
Here, we would like to contrast them with two other non-Markovianity measures. 
We use $\mathcal{N}_t$, which counts the number of time steps for which the EPR is negative. 
Further, we employ $\mathcal{N}_{\mathrm{C}}$ which counts how often the EPR crosses from being positive to being negative and vice versa. 
They are shown in Fig.~\ref{fig:app_measures}(c), (d).

We find that all four measures qualitatively yield the same observation, see Fig.~\ref{fig:app_measures}. 
The LLG equation always leads to a measure of zero for both configurations. 
The inertial LLG equation shows a more complex behaviour with the measures always being zero for the parallel configuration, but non-zero for the canted configuration. 
Lastly, the measure for the os-LLG equation is always non zero for both configurations, whereby the measure for the canted configuration is greater or equal to the measure for the parallel configuration. 
They are only equal for the $\mathcal{N}_t$ measure. 

\end{document}